\documentclass[11pt,number,sort&compress]{elsarticle}
\usepackage{color, amsmath, amssymb}
\usepackage{axodraw,epsf,float,booktabs,slashed}

\usepackage{vmargin}
\setpapersize{A4}
\setmarginsrb{2.2 cm}{2.2 cm}{2.2 cm}{2.2 cm}{0.0 cm}{0.0 cm}{0.0 cm}{1.2 cm}

\def\caja{\mathsurround=0pt}
\def\eqalign#1{\,\vcenter{\openup1\jot \caja
        \ialign{\strut \hfil$\displaystyle{##}$&$
         \displaystyle{{}##}$\hfil\crcr#1\crcr}}\,}
\numberwithin{equation}{section}

\renewcommand{\theequation}{\thesection.\arabic{equation}}
\def\equn{\refstepcounter{equation}\eqno({\rm \theequation})}

\def\npb#1#2#3{{\rm Nucl. Phys. B}{\bf \ #1}, #3 (#2)}
\def\cqg#1#2#3{{\rm Class. and Quant.\ Grav.} {\bf  #1}, #3 (#2)}
\def\hepth#1{[hep-th/#1]}
\def\hepph#1{[hep-ph/#1]}
\def\spa#1.#2{\left\langle#1\,#2\right\rangle}
\def\spb#1.#2{\left[#1\,#2\right]}
\def\eps{\epsilon}
\def\I{{\cal I}}
\def\la{\langle}
\def\ra{\rangle}
\def\Mloop{M^{\oneloop}}
\def\Mtree{M^{\tree}}

\def\Aloop{A^{\oneloop}}
\def\dlips{d{\text{LIPS}}}
\def\BR#1#2{[#1|{K_{abc}}|#2\ra}
\def\BRTTT#1#2{\la#1^+|\slashed{K}_{abc}|#2^+\ra}
\def\NeqEight{\Neq8}
\def\NeqSix{\Neq6}
\def\NeqFour{\Neq4}
\def\NeqOne{\Neq1}

\def\deff{d_{\text{eff}}}
\newcommand{\oneloop}{\text{one-loop}}
\newcommand{\tree}{\text{tree}}
\newcommand\ie{i.e.}
\newcommand{\Neq}[1]{\mathcal{N} = #1}
\DeclareMathOperator{\Perm}{\mathcal{P}}
\DeclareMathOperator{\Ord}{\mathcal{O}}
\DeclareMathOperator{\tr}{\mathrm{tr}}
\newcommand\trunc{\text{trunc}}
\newcommand\figref[1]{fig.~\ref{#1}}

\begin{document}

\title{\bf Perturbative expansion of ${\cal N} <8$
       Supergravity}

\author{David~C.~Dunbar}
\ead{d.c.dunbar@swan.ac.uk}

\author{James~H.~Ettle}
\ead{j.h.ettle@swan.ac.uk}

\author{Warren~B.~Perkins}
\ead{w.perkins@swan.ac.uk}

\address{Department of Physics,
  Swansea University,
  Swansea, SA2 8PP, UK}

\begin{abstract}
    We characterise the one-loop amplitudes for $\NeqSix$ and
    $\NeqFour$ supergravity in four dimensions.  For $\NeqSix$ we find
    that the one-loop $n$-point amplitudes can be expanded in terms of
    scalar box and triangle functions only.  This simplification is
    consistent with a loop momentum power count of $n-3$, which we
    would interpret as being $n+4$ for gravity with a further $-7$
    from the $\NeqSix$ superalgebra. For $\NeqFour$ we find that
    the amplitude is consistent with a
    loop momentum power count of $n$, which we would interpret as
    being $n+4$ for gravity with a further $-4$ from the $\NeqFour$
    superalgebra. Specifically the $\NeqFour$ amplitudes contain
    non-cut-constructible rational terms.
\end{abstract}

\maketitle

\section{Introduction}

Superficially the perturbative expansion of gravity scattering
amplitudes~\cite{Qgrav} is extremely complicated and power counting
suggests the theory is plagued with ultra-violet divergences.
However, there is growing evidence that the ultra-violet behaviour of
gravity theories is significantly softer than expected.  The bulk of
this evidence has arisen from studies of explicit on-shell scattering
amplitudes rather than formal structures.  The underlying drivers for
this behaviour remain unclear.  Even at tree level surprises have
recently been noted: the large momentum behaviour of tree scattering
amplitudes has a softer behaviour than expected~\cite{Bedford:2005yy,
  Cachazo:2005ca, BjerrumBohr:2005jr, Benincasa:2007qj} and a rich
structure of relationships between the tree amplitudes has been
uncovered~\cite{Bern:2008qj, BjerrumBohr:2010zb, BjerrumBohr:2010zs,
  Tye:2010kg, BjerrumBohr:2010yc, Vaman:2010ez, Elvang:2010kc,
  Feng:2010hd}, which go beyond the well known KLT
relations~\cite{Kawai:1985xq}.

At loop level, the softest theory is expected to be maximally
supersymmetric $\NeqEight$ supergravity~\cite{ExtendedSugra} .
Reexaminations of the perturbative expansion of $\NeqEight$ have
uncovered evidence that this theory has a softer UV structure than
previously thought~\cite{ConventionalSusy}.  Explicit calculations of
physical scattering amplitudes have shown that the four-graviton
amplitude is finite at two~\cite{BDDPR},
three~\cite{Bern:2007hh,Bern:2008pv} and four
loops~\cite{Bern:2009kd}.  In particular, the results indicate
cancellations between diagrams beyond these explicit in any known
formalism.  At one-loop $\NeqEight$ amplitudes for arbitrary numbers
of external gravitons have been shown to have a very restricted form,
to $O(\epsilon)$:
$$
A = \sum_i c_i I_4^i \equn
$$ where $I_4^i$ are scalar box-functions and $c_i$ are rational
coefficients~\cite{MaxCalcsA,MaxCalcsB,MaxCalcsC}.  This ``no-triangle
hypothesis''~\cite{BjerrumBohr:2006yw} must result from a much
stronger cancellation within supergravity theories than previously
thought and has been checked by explicit computations up to seven
points~\cite{MaxCalcsA,MaxCalcsB,MaxCalcsC,BjerrumBohr:2006yw} and
proven within a string-based rules formalism~\cite{BjerrumBohr:2008ji}
.  Both of these calculations indicate that in the UV limit the
behaviour of $\NeqEight$ supergravity tracks that of $\NeqFour$
super-Yang-Mills.  This opens the possibility that $\NeqEight$
supergravity is a finite quantum field theory of gravity.
Although potential counterterms may exist at
high loop order~\cite{NewCounterterms},   
there is no 
evidence contrary to finiteness at this point.

In~\cite{Bern:2007xj} and implicitly in~\cite{BjerrumBohr:2006yw} the
source of these cancellations was examined.  When calculating a
one-loop amplitude in a general gravity theory we sum over
diagrams. Let $m$ be the number of legs attached to the loop, $m\leq
n$.  We expect loop momentum integrals of the form
$$
I_m[ P^{2m}[\ell] ] \equn$$ where $P^{2m}[\ell]$ is a polynomial of
degree $2m$ in the loop momentum $\ell$. Cancellations between
diagrams can reduce the effective degree of the loop momentum
polynomial. We denote this effective degree by $\deff$.  The
traditional expectation within supergravity theories is that
cancellation between particle types within a supermultiplet reduces
the degree of the loop momentum polynomial from $2m$ to $\deff=2m-r$,
where $r$ depends upon the degree of supersymmetry. For maximal
supergravity $r=8$~\cite{GravityStringBasedA,GravityStringBasedB} is
manifest within the ``string-based rules'' method.  However the
no-triangle hypothesis indicates that further cancellations arise,
resulting in $\deff=m-4$.  This suggests a degree of $m + 4$ (rather
than $2m$) for pure gravity, reduced by $8$ by the $\Neq8$
supersymmetry.
In this article we explore the perturbative expansion of $\NeqSix$ and
$\NeqFour$ supergravity theories to examine their UV behaviour.  A
starting hypothesis for the reduction in the degree of the loop
momentum polynomial is
$$
\deff=(m+4)-r \equn\label{powerhypothesis}
$$
where $r=4$ for $\NeqFour$ supergravity and $r=6$ for $\NeqSix$
supergravity.  To understand the implications of this for the
structure of these amplitudes, we recall that a general one-loop
amplitude in a theory of massless particles can be expressed, after a
Passarino--Veltman reduction~\cite{PassVelt}, in the form
$$
\Aloop_n=\sum_{i\in \cal C}\, a_i\, I_4^{i} +\sum_{j\in \cal D}\,
b_{j}\, I_3^{j} +\sum_{k\in \cal E}\, c_{k} \, I_2^{k} +R_n\, ,
\equn\label{DecompBasis}
$$ 
where the $I_f$ are $f$-point scalar integral functions and the $a_i$
etc. are rational coefficients. $R_n$ is a purely rational term.  For
$\deff\geq n$ we expect this full generic form, while for $\deff<n$
the rational term is absent, for $\deff\leq n-3$ the bubbles $I_2$ are
also absent and for $\deff \leq n-4$ only the box functions appear.

For $\NeqSix$ our explicit calculations indicate $\deff=n-3$, \ie\
$r=7$.
Compared with \eqref{powerhypothesis} there is an extra reduction in
the power count by one for $\NeqSix$ amplitudes, giving them a
simplified expansion:
$$
\Mloop_n{}^{,\NeqSix}=\sum_{i\in \cal C}\, a_i\, I_4^{i} +\sum_{j\in
  \cal D}\, b_{j}\, I_3^{j}.  \equn\label{basisn6}
$$ 
This is consistent with the expectations of
\cite{Bern:2007xj,BjerrumBohr:2008ji}.  For $\NeqFour$ we find
amplitudes consistent with $\deff=n$, implying that $r=4$ and $R_n\neq
0$ in eq.~(\ref{DecompBasis}).  This contradicts previous
expectations~\cite{Bern:2007xj,BjerrumBohr:2008ji}.  The evidence for
this, together with a discussion of the implications, will form the
remainder of this article.

\section{IR consistency and Choice of Integral Function Basis}
For one-loop amplitudes IR consistency imposes a system of constraints
on the rational coefficients of the integral functions.  For the
matter multiplets~\cite{DunNorB} there are in fact no IR singular
terms in the amplitude, so the singular terms in the individual
integral functions cancel.  This gives enough information to fix the
coefficients of the one- and two-mass triangles in terms of the box
coefficients. The three-mass triangle is IR finite, so its coefficient
is not determined by these constraints.  It is convenient to combine
the boxes and triangles in such a way that these infinities are
manifestly absent.  There are several ways to do
this~\cite{BDDKa,BBDP,Britto:2005ha,Dunbar:2008zz}, here we choose to
work with truncated box functions
$$
I_4^{\trunc}= I_4 - \sum_i \alpha_i { \frac{(-s_i)^{-\eps}}{\eps^2} }
\equn
$$ 
where the $\alpha_i$ and $s_i$ are chosen to make $I_4^{\trunc}$ IR
finite.  This effectively incorporates the one- and two-mass triangles
together with the box integral functions.  Using these truncated
boxes, the coefficients of the one and two-mass triangles vanish and
the amplitudes can be written as
$$
\Mloop_n=\sum_{i\in \cal C}\, a_i\, I_4^{i,\trunc} +\sum_{j\in\cal D'}
b_{j}\, I_3^{j,\text{3-mass}} +\sum_{k\in \cal E}\, c_{k} \, I_2^{k}
+R_n\, , \equn\label{DecompBasisTruncated}
$$ 
with the single additional constraint $\sum c_k=0$.

\section{ $\NeqSix$ one-loop amplitudes}

At one-loop our $\NeqSix$ supergravity theory is specified by its
particle content and tree amplitudes.  There are two possible
multiplets: the vector multiplet and the matter multiplet, with
particle contents as follows:
\begin{center}
  \begin{tabular}{cccccccccc}
    \toprule
    Helicity  & \;\;$2$\;        &  \;\;$3/2$ \;   & 
    \;\;$1$ \; &   \;\;$1/2$ \;  &  \;\;$0$ \;  &  \;$-1/2$\;  & \;$-1$\;  & \;$-3/2$\;  & \;$-2$\; \cr \midrule
    vector   & $1$        & $6$   & $16$   & $26$   &   $30$  & $26$  &
    $16$ & $6$ & $1$ \\ 
    matter  &    $0$    & $1$   &  $6$  & $15$   & $20$   & $15$  & $6$ & $1$ & $0$ \cr\bottomrule
  \end{tabular}
\end{center}
The contributions to the one-loop $n$-graviton scattering amplitude
from the two $\NeqSix$ multiplets satisfy
$$
M^{\NeqSix,\text{vector}} = M^{\NeqEight} -2M^{\NeqSix,\text{matter}}.
\equn
$$
As $M^{\NeqEight}$ is known, it is sufficient to compute the
contribution from the matter multiplet alone.

\subsection{MHV amplitudes}
The one-loop $n$-point MHV amplitude\footnote{For clarity we suppress
  a factor of $i(\kappa/2)^{n-2}$ in each tree amplitude and
  $i{(\kappa/2)^n/(4\pi^2)}$ in each one-loop amplitude.} in $\NeqEight$
supergravity is~\cite{MaxCalcsB}
$$
\eqalign{ M_n^{\text{one-loop},\NeqEight} & (1^-, 2^-, 3^+, \ldots,
  n^+) = \cr & \frac{(-1)^n}{8} \, \spa1.2^8 \sum_{1 \leq a < b \leq n
    \atop M, N} h(a, M, b) h(b, N, a) \tr^2[a\, M\, b\, N]\,
  I_4^{aMbN}\ +\ \Ord(\eps)\,, \cr} \equn\label{MHVTree}
$$ 
where $h(a, M, b)$ are the ``half-soft'' functions of
ref.~\cite{MaxCalcsB} and $I_4^{aMbN}$ are the ``two-mass-easy''
scalar box functions with massless legs $a$ and $b$ and massive
clusters $M$ and $N$.  The summation includes the degenerate cases
where $M$ or $N$ reduce to a single massless leg. The half-soft
functions have the explicit form
$$
\eqalign{ h(a,\{1,2,\ldots,n\},b) &\equiv \frac{\spb1.2}{\spa1.2} { [3
    | {K_{12}} |a \ra [4 | {K_{123}} |a \ra \cdots [n| {K_{1\cdots
        n-1}} |a \ra \over \spa2.3\spa3.4 \cdots \spa{n-1,}.{n} \,
    \spa{a}.1 \spa{a}.2\spa{a}.3 \cdots \spa{a}.{n} \, \spa1.{b}
    \spa{n}.{b} } \cr &\hskip1cm + \Perm(2,3,\ldots,n), \cr}
\equn\label{NonRecursiveH}
$$
where we are using the usual spinor products $ \spa{j}.{l} \equiv
\langle j^- | l^+ \rangle = \bar{u}_-(k_j) u_+(k_l)$ and
$\spb{j}.{l}\equiv \langle j^+ | l^- \rangle = \bar{u}_+(k_j)
u_-(k_l)$, and where $\BR{i}{j}$ denotes $\BRTTT{i}{j}$ with
$K_{abc}^\mu =k_a^\mu+k_b^\mu+k_c^\mu$ and $s_{ab}=(k_a+k_b)^2$, etc.

The $\NeqSix$ matter multiplet's contribution to one-loop $n$-point
MHV amplitudes has vanishing three-mass triangle coefficients.  The
bubble coefficients also vanish as explicitly shown in
appendix~\ref{BubblesAppendix}.  Considering the rational terms,
$R_n$, the existence of an overall $\deff$ that ensures that the
bubble coefficients vanish would also ensure the vanishing of $R_n$.
Additionally, power counting in the string-based rules
\cite{GravityStringBasedA,GravityStringBasedB} gives $R_4=R_5=0$, and
if we assume $R_n$ could be recursively generated from $R_{n-1}$, this
would be sufficient to ensure $R_n=0$ for all $n$.

Consequently these contributions can be expressed purely as sums of
truncated boxes with a single negative helicity leg in each massive
corner, as shown in \figref{BoxFigure}.  The box coefficients may be
determined using unitarity methods~\cite{BDDKa} including quadruple
cuts~\cite{BrittoUnitarity}.  To use quadruple cuts we require the MHV
tree amplitudes for $n-2$ gravitons and a pair of particles of
helicity $\pm h$
$$
M( 1^- , 2^{-h} , 3^{+h} , 4^+ \cdots n^+) = \biggl(
\frac{\spa1.3}{\spa1.2} \biggr)^{2h-4} M( 1^- , 2^{-} , 3^{+} , 4^+
\cdots n^+) \equn
$$
where the MHV amplitudes from $n$-gravitons are given
in~\cite{BerGiKu}.  We find the box-coefficients are related to the
maximally supersymmetric case by simple factors, as in
QCD~\cite{BBDP},
\begin{multline}
  M_n^{\NeqSix,\text{matter}}(1^-, 2^-, 3^+, \ldots, n^+) = \\
  {(-1)^n\over 8} \spa1.2^8 \sum_{2 < a < b \leq n \atop 1 \in M, 2
    \in N} \left( -{ \spa{1}.{a}\spa{2}.{a}\spa{1}.{b}\spa{2}.{b} \over
      \spa{a}.b^2 \spa{1}.{2}^2 } \right) h(a, M, b) h(b, N, a)
  \tr^2[a\, M\, b\, N]\, \I_4^{aMbN,\trunc}
  \label{MHVN6}
\end{multline}
This gives an all-$n$ expression for the amplitude consistent with a
loop momentum power count of $n-3$ in agreement with previous results.

\begin{figure}[H]
  \begin{center}
    {
      \begin{picture}(250,144)(0,0)
        \SetOffset(125,72)
        \Line(-40,0)(0,40)
        \Line(0,40)(40,0)
        \Line(40,0)(0,-40)
        \Line(0,-40)(-40,0)
        \Line(0,40)(0,60)
        \Line(0,-40)(0,-60)
        \Text(0,62)[bc]{$b^+$}
        \Text(0,-62)[tc]{$a^+$}
        \Line(40,0)(60,0) \Line(40,0)(55,15) \Line(40,0)(55,-15)
        \Line(-40,0)(-60,0) \Line(-40,0)(-55,15) \Line(-40,0)(-55,-15)
        \Text(62,1)[lc]{$2^-$}
        \Text(-62,1)[rc]{$1^-$}
        \Vertex(53,8){0.5} \Vertex(54,4.5){0.5}
        \Vertex(53,-8){0.5} \Vertex(54,-4.5){0.5}
        \Vertex(-53,8){0.5} \Vertex(-54,4.5){0.5}
        \Vertex(-53,-8){0.5} \Vertex(-54,-4.5){0.5}
        \Text(-75,0)[rc]{$\displaystyle M \left\{\vphantom{\frac12}\right.$}
        \Text(75,0)[lc]{$\displaystyle \left.\vphantom{\frac12}\right\}N$}
      \end{picture}
    }
    \\
    \caption{ The box functions appearing in the $\NeqSix$ MHV
      one-loop amplitude\label{BoxFigure} }
  \end{center}
\end{figure}

\subsection{Six-point NMHV}
The six-point next-to-MHV (NMHV) amplitude contains several features
that are not present in the MHV amplitudes: in addition to the
one-mass truncated boxes the amplitude also contains two-mass-hard
truncated boxes and three-mass triangles.

\begin{figure}[H]
  \begin{center}
    \begin{picture}(150,100)(-50,0)
      \Line(30,30)(30,70)
      \Line(70,30)(70,70)
      \Line(30,30)(70,30)
      \Line(70,70)(30,70)
      \Line(30,30)(20,20)
      \Line(70,30)(80,20)
      \Line(30,70)(20,70)
      \Line(30,70)(20,80)
      \Line(30,70)(30,80)
      \Line(70,70)(80,80)
      \Text(13,14)[l]{$f^+$}
      \Text(78,14)[l]{$e^-$}
      \Text(7,72)[l]{$a^+$}
      \Text(12,87)[l]{$b^-$}
      \Text(26,88)[l]{$c^-$}
      \Text(83,89)[l]{$d^+$}
      \Text(-40,50)[l]{$I_4^{(abc)def}$}
      \SetWidth{1.5}
    \end{picture}
    \begin{picture}(150,100)(-50,0)
      \Line(30,30)(30,70)
      \Line(70,30)(70,70)
      \Line(30,30)(70,30)
      \Line(70,70)(30,70)
      \Line(30,30)(20,20)
      \Line(70,30)(80,20)
      \Line(30,70)(20,70)
      \Line(30,70)(20,80)
      \Line(30,70)(30,80)
      \Line(70,70)(80,80)
      \Text(13,14)[l]{$f^-$}
      \Text(78,14)[l]{$e^+$}
      \Text(7,72)[l]{$a^-$}
      \Text(12,87)[l]{$b^+$}
      \Text(26,88)[l]{$c^+$}
      \Text(83,89)[l]{$d^-$}
      \Text(-40,50)[l]{$I_4^{(abc)def}$}
      \SetWidth{1.5}
    \end{picture}
    \begin{picture}(150,100)(-50,0)
      \Line(30,30)(30,70)
      \Line(70,30)(70,70)
      \Line(30,30)(70,30)
      \Line(70,70)(30,70)
      \Line(30,30)(20,20)
      \Line(70,30)(80,20)
      \Line(30,70)(20,70)
      \Line(30,70)(30,80)
      \Line(70,70)(70,80)
      \Line(70,70)(80,70)
      \Text(13,12)[l]{$a^-$}
      \Text(78,12)[l]{$f^+$}
      \Text(12,72)[l]{$b^-$}
      \Text(26,89)[l]{$c^+$}
      \Text(66,90)[l]{$d^-$}
      \Text(83,72)[l]{$e^+$}
      \Text(-40,50)[l]{$I_4^{a(bc)(de)f}$}
    \end{picture}
    \begin{picture}(150,100)(-50,0)
      \Line(30,30)(70,30)
      \Line(30,30)(50,70)
      \Line(70,30)(50,70)
      \Line(30,30)(20,20)
      \Line(30,30)(20,40)
      \Line(70,30)(80,20)
      \Line(70,30)(80,40)
      \Line(50,70)(40,80)
      \Line(50,70)(60,80)
      \Text(13,12)[l]{$a^-$}
      \Text(8,42)[l]{$b^+$}
      \Text(26,85)[l]{$c^-$}
      \Text(66,85)[l]{$d^+$}
      \Text(83,42)[l]{$e^-$}
      \Text(78,12)[l]{$f^+$}
      \Text(-40,50)[l]{$I_3^{(ab)(cd)(ef)}$}
  \end{picture}
  \\
  \caption{ The box-functions appearing in the NMHV six-point one-loop
    amplitude\label{NMHVBoxFigure} } \end{center}
\end{figure}
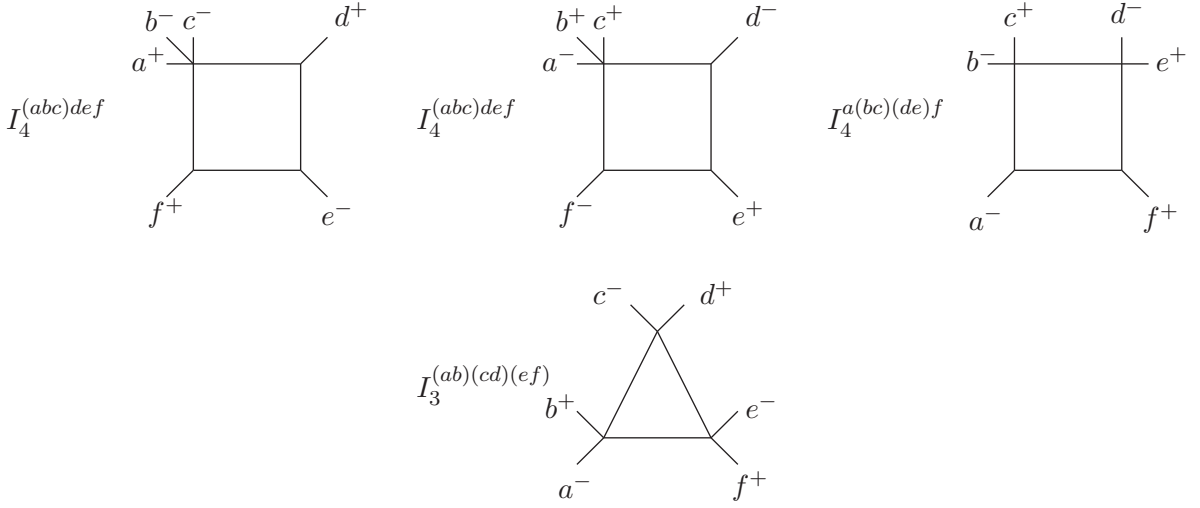

In terms of these integral functions the amplitude is,
$$
\eqalign{ &{\cal M}^{\NeqSix,matter}_6 (1^-,2^-,3^-,4^+,5^+,6^+)= \cr
  &
  \sum_{(abd)\in P_3(123) ;(cef)\in P_3(456)} c^{\,a(bc)(de)f}
  I_4^{a(bc)(de)f,\trunc}
  +\sum_{(adf)\in P_3(456);(bcd)\in P_3(123)} c^{\,(abc)def}_{\NeqSix}
  I_4^{(abc)def,\trunc} \cr & +\sum_{(adf)\in P_3(123);(bcd)\in
    P_3(456)} \overline{c}^{\,(abc)def}_{\NeqSix}
  I_4^{(abc)def,\trunc}
  +\sum_{(bde)\in P_3(456)} c^{(1b),(2d),(3e)}_{\NeqSix}
  I_3^{(1b)(2d)(3e)} \,.  \cr} \equn\label{boxsumEQ}
$$
The sums run over the permutations of indices $1, \dots ,6$, modulo
symmetries of the integral functions $I_4^{(abc)def}$ and
$I_4^{a(bc)(de)f}$.

The two-mass-hard box coefficients are
$$
\eqalign{ c^{\,a^-(b^-c^+)(d^-e^+)f^+}_{\NeqSix} = {i \over 2} {
    s_{bc}s_{de} s_{af}^2 (K_{abc}^2) \BR{a}d\BR{c}f\BR{c}d^6 \over
    \spb{a}.b\spb{b}.c^2 \spa{d}.e^2 \spa{e}.f \BR{a}{d}\BR{a}e
    \BR{b}e\BR{c}f \BR{a}f^2 } \cr}, \equn
$$
the one-mass box coefficients are
$$
\eqalign{
  & c^{\,(a^-b^+c^+)d^-e^+f^-}_{\NeqSix} = \cr {i \over 2} &
  {\spa{d}.{e}^2\spa{e}.{f}^2 \spb{d}.{e}\spb{e}.{f} \BR{e}{a}^6
    \Bigl( \spa{a}.b\spb{b}.c \BR{f}{c} \spb{d}.{a} + \spb{a}.{b}
    \spa{b}.{c}\spb{c}.{d}\BR{f}{a} \Bigr) \over
    \spa{a}.{b}\spa{b}.{c}\spa{a}.{c} \spb{d}.f^2 \BR{d}b\BR{f}b
    \BR{d}c\BR{f}c K^2_{abc} }\,\,, \cr} \equn
$$
and
$$
\overline{c}^{\,(a^+b^-c^-)d^+e^-f^+}_{\NeqSix} =
c^{\,(a^-b^+c^+)d^-e^+f^-}_{\NeqSix}|_{\lambda \leftrightarrow
  \bar\lambda}.  \equn$$

The three-mass triangle coefficients can be evaluated using analytic
techniques~\cite{Forde:2007mi,BjerrumBohr:2007vu,UsUnitarity} and are
$$
c^{\,(a^-b^+)(c^-d^+)(e^-f^+)}_{\NeqSix}=- {1 \over
  s_{ab}s_{cd}s_{ef} } \sum_{i=1}^6 C_{A_i} { \la B_6 |
  [K_{ab},K_{cd}]|A_i\ra \over 2 \la A_i | K_{ab} K_{cd} |A_i\ra },
\equn$$ where
$$
\{ |A_i\ra \} = \{ |b\ra , |f\ra , \;
K_{ef}K_{cd} | d \ra ,\;
K_{ab}K_{cd}|d\ra ,\;
K_{ef}K_{cd}|b\ra, \;
K_{ab}K_{cd}|f\ra \},
\equn$$
with
\begin{equation}
  \begin{aligned}
    |B_1\rangle = |B_2\rangle &= |X\rangle \\
  |B_3\rangle = |B_4\rangle &=   K_{ef}K_{cd} |Y\ra \\
    |B_5\rangle = |B_6\rangle &= K_{ab}K_{cd} |Z\ra 
  \end{aligned}
\end{equation}
where
$$
|X\rangle=  |a\ra [d|f|e\ra+ |e\ra [d|b|a\ra, \;
|Y\rangle = |c\ra
    [f|b|a\ra+ |a\ra [f|d|c\ra, ;\;
|Z\rangle =  |e\ra
    [b|d|c\ra+ |c\ra [b|f|e\ra,
\equn
$$
and
$$
C_{A_i} = { \prod_{j=1}^5 \la B_j | A_i \ra \over \prod_{j \neq i} \la
  A_j | A_i \ra }.  \equn$$

This explicit six-point amplitude has all the correct cuts and is,
again, consistent with a loop momentum power count of $n-3$.  The
absence of cut-constructible bubble terms can be seen from the
$\NeqSix$ version of the analysis in section (3.3) of
reference~\cite{BjerrumBohr:2006yw}.  We have presented results for
external gravitons: the box-coefficients for other external states may
be obtained using supersymmetric Ward identities~\cite{Bidder:2005in}.

\section{ $\NeqFour$ one-loop amplitudes}

The particle content multiplicities of $\Neq4$ graviton and matter
multiplets are as follows:
\begin{center}
  \begin{tabular}{cccccccccc}
    \toprule
    Helicity  & \;\;$2$\;        &  \;\;$3/2$ \;   & 
    \;\;$1$ \; &   \;\;$1/2$ \;  &  \;\;$0$ \;  &  \;$-1/2$\;  & \;$-1$\;  & \;$-3/2$\;  & \;$-2$\; \cr \midrule
    graviton   & $1$  & $4$ & $6$ & $4$ & $2$   & $4$ & $6$ & $4$ & $1$    \\ 
    matter  & $0$ & $0$ & $1$ & $4$ & $6$ & $4$ & $1$ & $0$ & $0$  \cr\bottomrule
  \end{tabular}
\end{center}
For convenience, we will calculate the one-loop amplitude using the
$\Neq4$ matter multiplet, which is related to the amplitude containing
the graviton by
$$
M^{\NeqFour,\text{graviton}} = M^{\NeqEight}
-4M^{\NeqSix,\text{matter}}+2M^{\NeqFour,\text{matter}}.  \equn
$$ 
To order $\epsilon^0$, the four-point one-loop $\NeqFour$ amplitude is
given by~\cite{GravityStringBasedB}
$$ 
\eqalign{ M^{\text{1-loop},\NeqFour}(1^-,2^-,3^+,4^+) &= {F \over
    2s^4} \bigg( {(t-u)s\ln(-t/-u)} -
  {tu\bigl(\ln^2(-t/-u)+\pi^2\bigr)} + { s^2} \bigg) \cr} \equn$$
where
$$
F=\biggl( {st \spa1.2^4 \over \spa1.2\spa2.3\spa3.4\spa4.1
} \biggr)^2 = {stu} {\Mtree(1^-,2^-,3^+,4^+)} \equn
$$
and $s\equiv s_{12}$, $ t\equiv s_{14}$ and $u\equiv s_{13}$, are the
usual Mandelstam variables.  In terms of integral functions this
result can be expressed as
$$
\eqalign{ M^{\text{1-loop},\NeqFour}(1^-,2^-,3^+,4^+) &= {F \over
    2s^4} \bigg( {(t-u)s( I_2(u)-I_2(t) )} + (tu)^2 I_4^{\trunc}(t,u)
  + {s^2} \bigg).  \cr} \equn$$ As we can see, this $\NeqFour$
amplitude contains a rational term
$$
R_4= {F \over 2s^2} =\frac{1}{2} \biggl( {t \spa1.2^4 \over
  \spa1.2\spa2.3\spa3.4\spa4.1 } \biggr)^2.  \equn$$ The presence of a
rational term indicates that the power count is at least 4 in this
case. Since higher-point amplitudes must reduce to the four-point
amplitude in soft and factorisation limits, it appears inevitable that
rational terms also appear in all $n$-point amplitudes, indicating
that the power count for $\Neq4$ supergravity and one loop is \emph{at
  least} $n$.

The $n$-point MHV amplitude is
\begin{multline}
  M_n^{\NeqFour} (1^-, 2^-, 3^+, \ldots, n^+) = \\
  {(-1)^n\over 8} \, \spa1.2^8 \sum_{2 < a < b \leq n \atop 1\in M, 2
    \in N} \left( -{ \spa{1}.{a}\spa{2}.{a}\spa{1}.{b}\spa{2}.{b} \over
      \spa{a}.b^2 \spa{1}.{2}^2 } \right)^2 h(a, M, b) h(b, N, a)
  \tr^2[a\, M\, b\, N]\, \I_4^{aMbN, \trunc} \ \\
  \quad+\sum_{1\in A,2\in B} c_2(1,A;2,B) I_2 (P_A^2) +R_n,\hfill
  \label{MHVboxes}\end{multline}
where the sets $A$ and $B$, contain at least one positive helicity
leg. The bubble coefficients $c_2(1,A;2,B)$ are derived and given
explicitly in the appendix.  Previously it has been
suggested~\cite{Bern:2007xj,BjerrumBohr:2008ji} that $R_n=0$ for
$\NeqFour$ amplitudes. Our analysis suggests otherwise: we have
explicitly seen that $R_4\neq 0$.  As a further check we have
evaluated $R_5$ at a specific kinematic point (given in the appendix)
using string-based rules for gravity~\cite{GravityStringBasedA,
  GravityStringBasedB, Dunbar:2010xk}.  At this kinematic point we
find
$$
R_5=
18856.6 + 37772.0 i
\equn
$$

An analytic expression which matches this is
$$
R_5=\left( \sum_{C_{345},P_{12}}     R_5^a \right)+ R_5^b  
\equn
$$
where
\begin{align}
R_5^a &=-\frac{1}{2} \spa1.2^4{  \spb3.4^2   \spb2.5\spa2.3\spa2.4
\over \spa3.4^2\spa2.5\spa3.5\spa4.5 }
\\
R_5^b &=
-\spa1.2^4{  
\spb3.4\spb3.5\spb4.5 \over \spa3.4\spa3.5\spa4.5 }
\end{align}
The six terms in the summation of $R_5^a$ cover interchanging the
negative helicity  legs $1$ and $2$ and a cyclic permutation  over the
positive helicity legs $3$, $4$ and $5$. ( $R_5^a$ is symmetric under
exchange of  legs $3$ and $4$.) 

This analytic form can be obtained by a study of the singularities,
both spurious and physical, in the amplitude~\cite{RealSoonNow}.

\section{Beyond one-loop}

All supergravity theories in $D=4$ are one- and two-loop finite since
there is no $R^3$ supersymmetric counterterm, but at three loops a
potential $R^4$ counterterm exists~\cite{Kallosh}.  Until recently is
was widely believed that all supergravity theories would generate this
counterterm at three loops~\cite{ConventionalSusy}.  (In higher
dimensions multiple possible $R^4$ terms exist: for $D=8,10$ the
dimensional ``lifts'' of $\NeqEight$, $\NeqSix$ and $\NeqFour$ have
different counterterm structures~\cite{Dunbar:1999nj}, but for $D=4$
there is a unique $R^4$ counterterm consistent with supersymmetry.)

\begin{figure}[h]
\begin{center}
{
\begin{picture}(200,50)
\SetOffset(45,0)
\SetScale{0.4}
\SetWidth{4}
\Line(0,100)(270,100)
\Line(0,20)(270,20)
\Line(60,60)(180,60)
\SetWidth{2}
\COval(80,60)(60,30)(0){Black}{Purple}
\COval(80,60)(30,15)(0){Black}{White}
\COval(200,60)(60,30)(0){Black}{Purple}
\SetColor{White}
\SetWidth{20}
\Line(140,0)(140,120)
\SetColor{Black}
\SetWidth{3}
\DashLine(140,0)(140,120){5}
\end{picture}
}
\\
\caption{ The three particle cut of a three-loop
  amplitude \label{ThreeLoopFigure} }
\end{center}
\end{figure}
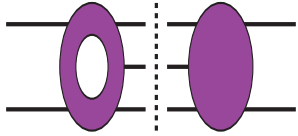

We can attempt to estimate the power counting of the multi-loop
amplitudes by considering various cuts.  In particular let us consider
the ``three-particle cut'' of the three-loop four-point amplitude,
$$
\int d\mathrm{LIPS}(l_i) \;M^{\oneloop} ( 1, 2,
\ell_1,\ell_2,\ell_3)\times M^{\tree} (3,4, \ell_1,\ell_2,\ell_3)
\equn
$$
as shown in \figref{ThreeLoopFigure}.

We can estimate the overall power counting by looking at the power
count of the uncut loop momenta. Note that we are looking at the
amplitude rather than individual diagrams.  For $\NeqEight$
supergravity, examining individual diagrams suggests the three-loop
power count is~\cite{BDDPR}
$$
\int d\ell_i { P_2( \ell_i,k_i ) \over \prod_{j=1}^{10}
  D_j(\ell_i,k_i) } \equn
$$ 
for a diagram with propagators $D_j$ and where $P_2$ is a polynomial
in the the loop momenta of degree 2.  However the
``no-triangle'' property suggest the power count in the indicated cut
is only $P_1$.  This gives a degree of divergence of
$$
3D-20+1 \equn
$$
making the amplitude divergent for
$$
D \geq 6.  \equn
$$
This is consistent with the explicit three-loop
computation~\cite{Bern:2008pv}.

For $\NeqSix$ and $\NeqFour$ (assuming the degree of divergence can be
inferred from this cut) we obtain
$$
\eqalign{ \text{$\NeqSix$:}\quad &3D-20+2 \cr \text{$\NeqFour$:}\quad
  &3D-20+5 \cr} \equn
$$
Both degrees of divergence are less than $-1$ for $D=4$ and so we
would predict that both theories remain finite at three-loops.  These
estimates must be taken with some caution: estimates of the power
counting in supergravity theories have proven wrong on many
occasions. Specifically, we cannot exclude further cancellations
within integrands and we are not sensitive to all possible terms.
Experience suggests that explicit calculations are required.

\section{Conclusions}

Explicit calculations of scattering amplitudes in $\NeqSix$ and
$\NeqFour$ supergravity theories indicate loop momentum power counts
of $n-3 \;(=n+4\;-7)$ and $n\; (=n+4\;-4)$ respectively. While the
former is in agreement with previous expectations, the latter in
not. In particular, the $\NeqFour$ amplitudes contain purely rational
terms.  We expect both these theories to remain finite up to
three-loops.

This research was supported by the STFC of the UK.

\appendix

\section{Bubbles in Supergravity MHV amplitudes}
\label{BubblesAppendix}

Here we present the bubble contributions to MHV amplitudes. Consider a
cut in the momenta $P=k_a+\cdots k_b$. The coefficient of the bubble
integral function $I_2(P^2)$ can be obtained from the cut,
$$
C_{a,\ldots,b} \equiv { i \over 2} \sum_h \int \dlips\biggl[ {\cal
  M}^{\tree}(-\ell_1^h,a,a+1,\ldots, b,\ell_2^{-h}) \times {\cal
  M}^{\tree}(-\ell_2^h,b+1,b+2,\ldots,a-1,\ell_1^{-h} ) \biggr] \, ,
\equn
$$
where $\int \dlips$ denotes integration over the on-shell phase
space of the $\ell_i$.  We must sum over the states in the $\NeqFour$
matter multiplet.  This cut vanishes unless we have a single negative
helicity leg and at least one positive helicity leg on each side.

There are a variety of techniques available to determine the bubble
coefficient from the cut: we will use the method of canonical
forms~\cite{UsUnitarity}.  We decompose the product of tree amplitudes
appearing in a two-particle cut in terms of canonical forms ${\cal
  F}_i$,
$$ 
\sum M^{\tree}(-\ell_1, \cdots, \ell_2) \times
M^{\tree}(-\ell_2,\cdots,
 \ell_1) = \sum_i c_i {\cal F}_i ({\ell_j}),
\equn
$$
where the $c_i$ are coefficients independent of $\ell_j$.  We then use
substitution rules to replace the ${\cal F}_i ({\ell_j})$ by the
evaluated forms $F_i(P)$ and obtain a bubble coefficient
$$
\sum_i c_i F_i(P) \equn
$$
For example, the simplest canonical form we use is
$$
{\cal H} _1(A ; B ; \ell) \equiv { \spa{\ell}.{B} \over \spa{\ell}.{A}
} \equn$$ which, for $\ell=\ell_1$ or $\ell_2$, evaluates to a
contribution to the bubble coefficient of
$$
H_1[A ; B ; P ] ={[A|P|B \ra \over [A|P|A\ra}.  \equn
$$
It is convenient to define extensions,
$$
{\cal H}_n ( A_i ; B_j ; \ell ) = { \prod_{j=1}^n \spa{B_j}.{\ell}
  \over \prod_{i=1}^n \spa{A_i}.\ell } \longrightarrow H_n[ A_i ; B_j
; P ]= \sum_i { \prod_{j=2}^n \spa{B_j}.{A_i} \over \prod_{j\neq i}
  \spa{A_j}.{A_i} } {\la B_1 | P | A_i ] \over \la A_i | P | A_i ] }
\; , \;\;\; \spa{A_i}.{A_j} \neq 0 .  \equn
$$
We will also need the special cases where $A_1=A_2=A$,
$$
{\cal H}_2^x( A, A ; B_1 ,B_2 ; \ell_i
)={\spa{B_1}.{\ell_1}\spa{B_2}.{\ell_2}\over
  \spa{A}.{\ell_1}\spa{A}.{\ell_2} } \longrightarrow { H}_2^x[ A, A ;
B_1 ,B_2 ; P ]= { [A|P|B_1\ra [A|P|B_2\ra\over [A|P|A\ra^2 }.  \equn
$$
and
$$
\eqalign{ {\cal H}^x_{2,1} &= {\spa{B_1}.{\ell_1} \spa{B_2}.{\ell_2}
    \spa{B_3}.{\ell_2} \over \spa{A}.{\ell_1} \spa{A}.{\ell_2}
    \spa{A_3}.{\ell_2} } \longrightarrow \left( { \spa{B_3}.{A} \over
      \spa{A_3}.{A} } H_2^x[A,A; B_1,B_2; P] + { \spa{B_3}.{A_3} \over
      \spa{A}.{A_3}} H_2[ A, A_3 ; B_1 , B_2 ; P]\right) \cr} \equn$$

We now return to the cut $C_{a\cdots b}$.  To be non-zero the set
$a\cdots b $ must contain exactly one negative helicity graviton and
at least one positive helicity graviton, i.e. must be of the form
$\{a_1^+,a_2^+,\cdots, a_{n_L}^+,m_1^-\}$ and the legs on the other
side must be $\{b_1^+,b_2^+,\cdots, b_{n_R}^+,m_2^-\}$.  The product
of tree amplitudes is then just a product of the two MHV trees.

When summing over the states in the multiplet, each tree amplitude is
proportional to the tree amplitude with two scalars up to a simple
factor. Summing over the tree amplitudes then yields
$$
\eqalign{ \sum_h \biggl[ {\cal M}^{\tree}(-\ell_1^h, a_1^+,a_2^+,\cdots
  a_{n_L}^+, m^-_1,\ell_2^{-h}) \times {\cal
    M}^{\tree}(-\ell_2^h,b_1^+,b_2^+,\cdots b_{n_R}^+,
  m_2^-,\ell_1^{-h} ) \cr = {\cal M}^{\tree}(-\ell_1^s,
  a_1^+,a_2^+,\cdots a_{n_L}^+, m^-_1, \ell_2^{s}) \times {\cal
    M}^{\tree}(-\ell_2^s,b_1^+,b_2^+,\cdots b_{n_R}^+, m_2^-,\ell_1^{s}
  ) ] \times \rho \cr} \equn
$$
Where the $\rho$-factor is
$$
\rho = \left( { \spa{m_1}.{m_2}^2 \spa{l_1}.{l_2}^2 \over
    \spa{m_1}.{l_1} \spa{m_1}.{l_2} \spa{m_2}.{l_1} \spa{m_2}.{l_2} }
\right)^A \equn$$ where $A=2$ for $\NeqFour$ and $A=3$ for $\NeqSix$
($A=1$ for a $\NeqOne$ matter multiplet).

Next, we rewrite the standard form of the MHV tree
amplitude~\cite{BerGiKu} so that the permutation is on the positive
helicity gravitons,
$$
\eqalign{ M_n^{\tree} &(1^s,2^+,3^+,\cdots,(n-2)^+,(n-1)^- ,n^s)
  \;=\;-i\spa{1}.{n-1}^4\spa{n}.{n-1}^4\times \cr & \biggl[{
    \spb{1}.2\spb{n-2}.{n-1} \over \spa{1}.{n-1} N(n) } \Bigl(
  \prod_{i=1}^{n-3} \prod_{j=i+2}^{n-1} \spa{i}.j \Bigr)
  \prod_{p=3}^{n-3} (-[p|K_{p+1\cdots n-1}|n\ra) +
  \Perm(2,3,\cdots,n-2) \biggr]\,, \cr} \equn\label{BGKform}
$$
where $N(n)=\prod_{i<j} \spa{i}.{j}$.  Labelling the negative helicity
leg $n-1$ as $m$ and legs $2$ to $n-2$ as $a_1\cdots a_{n'}$ and
identifying legs $1$ and $n$ with $\ell_1$ and $\ell_2$ gives
$$
\eqalign{ M_n^{\rm tree} &(l_1^s,a_1^+,a_2^+,\cdots,a_{n'}^+,m^-
  ,l_2^s) =-i\spa{l_1}.{m}^4\spa{l_2}.{m}^4
  \biggl[{ \spb{l_1}.{a_1}\spb{{a_{n'}}}.{m} \spa{l_1}.m \over
    \spa{l_1}.{m}^2 \spa{l_2}.{m}N_{n'} ( \prod_i
    \spa{l_1}.{a_i}\spa{l_2}.{a_i}\spa{a_i}.{m} \spa{l_1}.{l_2} } \cr
  \times & \left( \prod_{j=2}^{n'} \spa{l_1}.{a_j} \right) \left(
    \prod_{j=1}^{n'-1} \spa{a_j}.m \right) \left( \prod_{i=1}^{n'-1}
    \prod_{j=i+2}^{n'} \spa{a_i}.{a_j} \right) \prod_{p=2}^{n'-1}
  (-[a_p|\tilde K_{p}|l_2\ra)
  +\Perm(a_1,a_2,\cdots,a_{n'}) \biggr]\, \cr =-i &
  \spa{l_1}.{m}^3\spa{l_2}.{m}^3
  \biggl[ { \spb{{a_{n'}}}.{m} \spb{l_1}.{a_1} ( \prod_{i=1}^{n'-1}
    \prod_{j=i+2}^{n'} \spa{a_i}.{a_j}) \prod_{p=2}^{n'-1}
    (-[a_p|\tilde K_{p+1}|l_2\ra \over N_{n'} \spa{a_{n'}}.{m}
    \spa{l_1}.{l_2} \spa{l_1}.{a_1} ( \prod_{i=1}^{n'}
    \spa{l_2}.{a_i}) } + \Perm(a_1,a_2,\cdots,a_{n'}) \biggr] \cr}
\equn\label{BGKform2}
$$
where $\tilde K_p =k_{a_p}+\cdots k_{a_{n'}}+k_m$ and
$N_{n'}=\prod_{i<j} \spa{a_i}.{a_j}$.

Counting each factor of the form $\spa{A}.{l_i}$ or $\spb{A}.{l_i}$ as
having a loop-momentum weight of $+\frac{1}{2}$, the power count on
the cut momenta of a tree amplitude is of order $+2$.  For the
$\NeqSix$ multiplet, the $\rho$ factor contributes $-6$ so the cut is
of order $\ell_i^{-2}$ and thus~\cite{UsUnitarity} gives a bubble
coefficient of zero.

For the $\NeqFour$ matter multiplet, the cut is
$$
\sum_h {\cal M}^{\tree}(-\ell_1^h,\ldots, ,\ell_2^{-h}) \times {\cal
  M}^{\tree}(-\ell_2^h,\ldots,\ell_1^{-h} )= { \spa{m_1}.{m_2}^4 }
\sum_{P_l(a_i)}\sum_{P_r(b_i)} T_{(P_l;P_r)} \equn$$ where
$$
\eqalign{ T_{(P_l;P_r)} &= C_{P_l}C_{P_r} { \spa{l_1}.{l_2}^2
    \spb{l_1}.{a_1} \spb{l_1}.{b_1} \spa{m_1}.{l_1} \spa{m_1}.{l_2}
    \spa{m_2}.{l_1} \spa{m_2}.{l_2} \prod_{l=2}^{n_L-1}
    \spa{A_l}.{l_2} \prod_{r=2}^{n_R-1} \spa{B_r}.{l_2} \over
    \spa{l_1}.{a_1}\spa{l_1}.{b_1} \prod_{x\in \{ a_i,b_j\}}
    \spa{x}.{l_2} } \cr &= C_{P_l}C_{P_r} { \spa{m_1}.{l_1}
    \spa{m_2}.{l_1} [a_1|P|l_2\ra [b_1|P|l_2\ra \spa{m_1}.{l_2}
    \spa{m_2}.{l_2} \prod_{l=2}^{n_L-1} \spa{A_l}.{l_2}
    \prod_{r=2}^{n_R-1} \spa{B_r}.{l_2} \over
    \spa{l_1}.{a_1}\spa{l_1}.{a_2} \prod_{x\in \{ a_i,b_j\} }
    \spa{x}.{l_2} } \cr &= C_{P_l}C_{P_r} { \spa{m_1}.{l_1}
    \spa{m_2}.{l_1} \prod_{i=1}^{n_L} \spa{A_i}.{l_2}
    \prod_{j=1}^{n_R} \spa{B_j}.{l_2} \over
    \spa{l_1}.{a_1}\spa{l_1}.{b_1} \prod_{x\in \{ a_i,b_j\} }
    \spa{x}.{l_2} } \cr} \equn$$ and
$$
\eqalign{ | A_i\ra =\biggl\{\begin{matrix}
    \tilde K_{i}|a_i]  \;  & i \leq n_L-1   \\
    |m_1\ra \; & i=n_L
  \end{matrix}
  \;\;\;\ | B_j\ra =\biggl\{\begin{matrix}
    \tilde K_{j}'|b_j]  \;  & j \leq n_R-1   \\
    |m_2\ra \; & j=n_R
  \end{matrix}
  \cr} \equn
$$
$$
C_{P_L}= {( \prod_{i=1}^{n_L-1} \prod_{j=i+2}^{n_L} \spa{a_i}.{a_j})
  \over N(n_L) \spa{n_L}.{m_1} } ={ 1 \over \spa{n_L}.{m_1}
  \prod_{i=1}^{n_L-1} \spa{a_i}.{a_{i+1}} } \equn$$ We can rearrange
the $\ell_2$ dependant part,
$$
\eqalign{ &{ \spa{m_1}.{l_1} \spa{m_2}.{l_1} \prod_{i=1}^{n_L}
    \spa{A_i}.{l_2} \prod_{j=1}^{n_R} \spa{B_j}.{l_2} \over
    \spa{l_1}.{a_1}\spa{l_1}.{b_1} \prod_{x\in \{ a_i,b_j\} }
    \spa{x}.{l_2} }
  =
  \sum_ {x\in \{ a_i,b_j\} } D_{x} { \spa{m_1}.{l_1} \spa{m_2}.{l_1}
    \spa{m_1}.{l_2} \over \spa{l_1}.{a_1}\spa{l_1}.{b_1} \spa{x}.{l_2}
  } \cr} \equn\label{Eqcutstuff}
$$
with
$$
D_{x} = { \spa{m_2}.{x} \prod_{l=1}^{n_L-1} [ a_l|\tilde K_{l+1}|x\ra
  \prod_{k=1}^{n_R-1} [ b_k|\tilde K_{k+1}'|x\ra \over \prod_{y \neq
    x} \spa{x}.{y} } = { \prod_{l=1}^{n_L-1} [ a_l|\tilde K_{l+1}|x\ra
  \prod_{k=1}^{n_R} [ b_k|\tilde K_{k+1}' |x\ra \over
  \spb{b_{n_R}}.{m_2} \prod_{y \neq x} \spa{x}.{y} } \equn$$

Now for $x\neq a_1,b_1$ the term in eq.~(\ref{Eqcutstuff}) just gives
$H_3$ canonical forms. For $x=a_1$ or $b_1$ we get $H^x_{2,1}$ terms,
Putting the pieces together, we have a bubble coefficient of
$$
\eqalign{ c(m_1,\{a_i\} & ;m_2,\{b_i\} )= \spa{m_1}.{m_2}^4 \sum_{
    P_L,P_R } C_{P_L} C_{P_R} \biggl( \sum_{x\neq a_1,b_1} D_x H_3(
  x,a_1,b_1; m_1,m_2,m_1 ; P) \cr + & D_{a_1} H^{x}_{2,1}( a_1 , a_1 ,
  b_1 ; m_1,m_2,m_1 ; P) + D_{b_1} H^{x}_{2,1}( b_1 , b_1 , a_1 ;
  m_1,m_2,m_1 ; P) \biggr) \cr} \equn$$

\section{Kinematic Point}

\def\cent{\mu^{1/2}}

We use a kinematic point defined in terms of the following spinors:
\begin{align*}
  \lambda^{(1)}_\alpha &= \cent \begin{pmatrix} 46 + i \\ 14 + 18
    i \end{pmatrix},
  \\
  \lambda^{(2)}_\alpha &=  \cent \begin{pmatrix} 54+39i \\
    39+53i \end{pmatrix},
  \\
  \lambda^{(3)}_\alpha &= \cent \begin{pmatrix} 9+46i \\ 16+13i
  \end{pmatrix},
  \\
  \lambda^{(4)}_\alpha &= \cent
  \sqrt{\frac{42\,331}{4\,181\,993}} \begin{pmatrix}540 + 480i \\ 200
    + 170i
  \end{pmatrix},
  \\
  \lambda^{(5)}_\alpha &= \cent \begin{pmatrix}
    \sqrt{\frac{14\,499\,838\,743}{4\,181\,993}} \\
    (5\,099\,005\,787 +
    2\,200\,443\,816i)\sqrt{\frac{3}{20\,212\,741\,374\,784\,933}}
  \end{pmatrix},
\end{align*}
and the conjugate spinors $\tilde\lambda^{(i)}$ are given by
\begin{equation*}
  \tilde\lambda^{(i)}_{\dot\alpha} = \begin{cases}(\lambda^{(i)}_\alpha)^* &
    \text{for $i=1,2,3$,} \\
    -(\lambda^{(i)}_\alpha)^* & \text{for $i=4,5$.}
  \end{cases}
\end{equation*}
The numerical complexity of this point comes from the requirements
that it is real in Minkowski space and free from any
coplanarities. Momenta $4$ and $5$ have negative energy.
Additionally, we set the renormalisation scale $\mu^2 = 10^{-4}$.

\end{document}